\documentstyle[preprint,aps]{revtex}
\begin{document}
\draft

\flushbottom
\preprint{\vbox{
\hbox{IFT-P.019/94}
\hbox{April 1995}
}}
\title{Extra dimensions and color confinement}
\author{Vicente Pleitez }
\address{
Instituto de F\'\i sica Te\'orica\\
Universidade Estadual Paulista\\
Rua Pamplona, 145\\
01405-900-- S\~ao Paulo, SP\\
Brazil}
\maketitle
\begin{abstract}
We consider an extension of the ordinary four dimensional Minkowski space
by introducing additional dimensions which have their own Lorentz
transformation. Particles can transform in a different way under each
Lorentz group. We show that only quark interactions are slightly
modified and that color confinement is automatic since
these degrees of freedom run only in the extra dimensions. No
compactification of the extra dimensions is needed.
\end{abstract}
\narrowtext

It is very well known that in relativistic field theories all massless
particles have the same speed in vacuum: the speed of light $c$.
However, if the true space-time were direct product of several
four dimensional Minkowski space-time manifolds the Lorentz symmetry would
be $L\otimes L'\otimes L''\otimes\cdots$, with each factor $L,L',L''...$
having different limit velocity, say $c,c',c'',...$ respectively.

In the following, we will consider only two of such factors, say
$L\otimes L'$. The space-time is then, in this case, an eight dimensional
manifold. However, the four dimensions $x^\mu$, which we will call the
$x$-world, transform under $L\otimes L'$ like $({\bf4},{\bf1})$, while
$x^{\mu'}$ (the $x'$-world) transforms as $({\bf1},{\bf4})$. Here,
$\mu,\mu'=0,1,2,3$.

Now, let us introduce fields. Under $L\otimes L'$ the photon is assumed to
transform as $( {\bf4},{\bf1} )$, while gluons
transform as $( {\bf1},{\bf4} ) $. We can write the usual tensors
$F_{\mu\nu}$ for photons in the $x$-world and $F_{\mu'\nu'}^a$ for gluons
($a$ is the color index) in the $x'$-world.

Next, we will assume that two type of spinor fields do exist. One type
transforms as a Dirac spinor under $L$ but as a singlet under $L'$
and the other one transforms as Dirac spinor under both factors. The
former ones will be identified with leptons and the later ones with
quarks. For leptons the Dirac equation is as usual
in the $x$-world. We denote quark fields as $q_{\alpha\alpha'}(x,x')$,
where $\alpha$, $\alpha'$ represent spinor indices in the $x-$ and $x'-$worlds,
 respectively. Thus, the Dirac equation must be satisfied in each index
$\alpha$ and ${\alpha}'$ separately
\begin{equation}
\left(i\hbar c\gamma^\mu\partial_\mu-mc^2 \right)_{\alpha\beta}
\left( i\hbar {c}'\gamma^{\mu'}\partial_{\mu'}-m{c'}^2\right)_{\alpha'\beta'}
q_{\beta\beta'}(x,x')=0.
\label{qdirac}
\end{equation}
where $\alpha,\beta,\alpha',\beta'$ denote, in an obvious notation, spinor
indices. The free Lagrangian density reads then
\begin{equation}
{\cal L}_q(x,x')=\bar{q}_{\alpha\alpha'}(x,x')
\left(i\hbar c\gamma^\mu\partial_\mu-mc^2 \right)_{\alpha\beta}
\left( i\hbar {c}'\gamma^{\mu'}\partial_{\mu'}-m{c'}^2\right)_{\alpha'\beta'}
q_{\beta\beta'}(x,x')
\label{diraclag}
\end{equation}
Notice that $q_{\alpha\alpha'}(x,x')$ is a double-spinor since it transforms
as a spinor under each factor of $L\otimes L'$. It is different from the usual
 second rank spinor. The latter one has two spinor indices with respect to the
same Lorentz transformation. Notice that in general $q_{\alpha\alpha'}(x,x')
\not=q_{\alpha}(x)q_{\alpha'}(x')$.

Since gluons run only over the $x'$ world, this implies that color confinement
 is automatic in this context. It seems at first sight that in our scheme
there is not, besides the color confinement, any other observable effect. This
 is not the case but, in order to treat this issue it is
necessary to consider interactions.

With all those fields defined above, we can build interactions by introducing
covariant
derivatives. For leptons we have the usual quantum electrodynamics (QED).
In fact, we can assume that the
intermediate vector bosons $W^\pm$ and $Z^0$ transform like
the photon, and the Higgs boson is an scalar field under both
kind of Lorentz transformations. Hence, the standard electroweak
theory~\cite{wsg} remains the same for the lepton case.
On the other hand,
for quarks we obtain the gluon-photon-quark interactions which are
different, in general, from those of QED and quantum chromodynamics
(QCD)~\cite{qcd}. However, the deviations from pure QED and QCD are, in
principle, calculable.

{}From Eq.~(\ref{diraclag}) quark interactions arise in the usual
way if we define the covariant derivatives as
$\partial_\mu\to D_\mu=\partial_\mu-ieA_\mu$ and
$\partial_{\mu'}\to D^a_{\mu'}=\partial_{\mu'}+ig\,G^a_{\mu'}\lambda^a/2$,
with $\lambda^a,\;a=1,...8$
the $SU(3)$ Gell-Mann matrices. We also could have started with
a theory which is invariant under $SU(3)_c\otimes SU(2)_L\otimes U(1)_Y$, i.e.,
$\partial_\mu\to D^f_\mu$ and $\partial_{\mu'}\to D^a_{\mu'}$,
where $D^f_\mu$ is the $SU(2)\otimes U(1)$ covariant derivative
for the flavor $f$, but here we only will consider the first case.
As we said before, $q_{\alpha\alpha'}$ is a general double-spinor under
$L\otimes L'$. We have three possibilities for introducing interactions in the
 lagrangian of Eq.~(\ref{diraclag}).

In the first one, the covariant derivative (Abelian like QED or non-Abelian
like the electroweak interactions) is introduced only in the $x$-world part of
the lagrangian being, the $x'$-world part of quark fields an spectator.
For an Abelian QED-like gauge symmetry, the interaction lagrangian density is
\begin{equation}
L^{QED}_q=e\,\int\,d^3xd^3x'\,\bar{q}_{\alpha\alpha'}(x,x')\gamma^\mu_{\alpha
\beta}A_\mu(x,x') q_{\beta\alpha'}(x,x').
\label{sqed}
\end{equation}
We see that concerning the indices related to the $x$-world, Eq.~(\ref{sqed})
is in fact the usual QED interaction lagrangian density.

In the second case, it is the $x$ part of the quark fields which acts as an
spectator, while the QCD-like covariant derivative is introduced in the
$x'$-world part. Thus, the interaction reads
\begin{equation}
L^{QCD}_q=g\,\int\, d^3xd^3x'\,\bar{q}_{\alpha\alpha'}(x,x')\gamma^{\mu'}_
{\alpha'\beta'}
A^a_{\mu'}(x,x')\frac{\lambda^a}{2}q_{\alpha\beta'}(x,x'),
\label{sqcd}
\end{equation}
where we have omitted the color index in the quark fields. In this case,
concerning the indices running over the $x'$-world we obtain the QCD
interaction lagrangian.

Finally, we have both possibilities at the same time.
Then,  interactions are
\begin{equation}
L^{new}_q=
eg\,\int\;d^3xd^3x'\,\bar{q}_{\alpha\alpha'}\gamma^\mu_{\alpha\beta}A_\mu
\gamma^{\mu'}_{\alpha'\beta'}A^a_{\mu'}\frac{\lambda^a}{2}q_{\beta\beta'}.
\label{yuca}
\end{equation}
Notice that, since the gluon fields transform as singlet under $L$, from the
 point of view of an observer in the $x$-world the interaction in
Eq.~(\ref{yuca}) is of the form
\begin{equation}
 L^{new}_q=e\,\int\,d^3x\,\bar q(x)\gamma^\mu q(x)A_\mu(x)\,\phi(x),
\label{puxa}
\end{equation}
where $\phi(x)$ is the gluon fields viewed from the $x$-world. So, the
 interaction in Eq.~(\ref{puxa}) is of the non-renormalizable type in the
$x$-world.

Next, for simplicity, let us assume that $q_{\alpha\alpha'}(x,x')=
q_\alpha(x)q_{\alpha'}(x')$ in Eqs.~(\ref{sqed}), (\ref{sqcd}) and
(\ref{yuca}). Hence, in this simplified situation both $x$- and $x'$- world
 decouple from each other even in the quark sector. The interaction in
Eq.~(\ref{sqed}) becomes
\begin{equation}
L^{QED}_q=e\,\hbar c\int d^3x\,\bar{q}(x)\gamma^\mu q(x) A_\mu(x)\cdot\,W_G,
\label{qed}
\end{equation}
where we have defined
\begin{equation}
W_G=\int d^3x'\bar{q}(x')(i\hbar c'\gamma^{\mu'}\partial_{\mu'}-m{c'}^2)q(x').
\label{def1}
\end{equation}
On the other hand, the interactions Eq.~(\ref{sqcd}) is now
\begin{equation}
 L^{QCD}_q=g\,\hbar c'\int d^3x'\,\bar{q}_i(x')\gamma^{\mu'}
\frac{\lambda_{ij}^a}{2} q_j(x')A^a_{\mu'}\cdot\,W_A,
\label{qcd}
\end{equation}
with
\begin{equation}
W_A=\int d^3x\bar{q}(x)(i\hbar c\gamma^{\mu}\partial_{\mu}-m{c}^2)q(x).
\label{def2}
\end{equation}

Finally, the interactions in Eq.~(\ref{yuca}) is written as
\begin{equation}
L^{new}_{q}=\left[e\hbar c\int d^3x\,\bar{q}(x)\gamma^\mu q(x) A_\mu(x)\right]
_{em}\cdot
\left[g\hbar c'\int d^3x'\,\bar{q}_i(x')\gamma^{\mu'}\frac{\lambda_{ij}^a}{2}
 q_j(x')A^a_{\mu'}\right]_{color}.
\label{sim}
\end{equation}

Defining $\alpha'_s\equiv g^2/\hbar c'$ and $\alpha_s\equiv g^2/\hbar c$
we have
\begin{equation}
\alpha'_s=\alpha_s \,\cdot\frac{c}{c'}.
\label{gc}
\end{equation}
Thus, i) if $c'< c$, it means that $\alpha'_s>\alpha_s$ or, ii) if
$c'>c$, that $\alpha'_s<\alpha_s$. Of course the possibility $c'=c$
is not excluded but we will argue below that an interesting possibility is
the first one.

In QCD hadrons are considered as being bound states of
permanently confined quarks. However, until now, there is no proof
neither of the existence of bound states nor of the confinement
hypothesis in realistic $3+1$ theories.

The confinement of quarks and gluons is supposed to be explained, in
principle, in the context of QCD. Confinement
means that we cannot isolate or produce particles carrying
color. On the
other hand, no fractional electric charge has also been found until
now~\cite{pfs}.
Hence, since quarks carry out this sort of charge we can ask ourselves
if the mechanism for the color confinement and charge screening are
the same or not. It could be that quarks screen their color by vacuum
polarization effects, but the Coulomb fields which give us
information about their fractional electric charges remain unshielded.
There is no answer for this question based on ``first
principles'' but it was pointed out some years ago that in the
massive Schwinger model the screening of the electric charge is not
necessarily an indication for the color charges confinement~\cite{jas}. In
other words, it means that the absence of colored states cannot be
interpreted necessarily as a result of confinement in the usual
sense, and colorless quarks are possible.

In the context we have considered above, it is possible that, instead
of being the color confinement a dynamical consequence of the theory,
it should be an implication of extra space-time  dimensions.
Hence, we do not understand the neutralization of the color
degrees of freedom as the formation of hadrons but, as an effect due to
the fact that these degrees of freedom run out in extra space-time dimensions
in
which the speed limit is $c'$. We do not known the actual value of $c'$ but
only that it could be different from the usual speed
of light in vacuum $c$.

Usually, in the quark picture, decays or reactions involving hadrons
show symbolically the flow of quantum numbers during the process
under consideration. However, this is of limited use when dynamical
quantities are treated since we do not know how to represent the quark
confinement in the context of a theory of strong interactions, say QCD.
Hence, in the diagram representing the decays or reactions involving hadrons
our ignorance is contained within the usual ``shaded circles''
(form factors).

According to the uncertainty principle, a fluctuation
$\Delta E$ in the energy of a particle is not observable if it occurs in
a sufficiently
small time interval $\Delta t$. The maximum distance that a particle can
travel in this time is (we assume that the value of $\hbar$ is the same in
both $x$ and $x'$-worlds)
\begin{equation}
R\approx c\Delta t=\frac{\hbar c}{\Delta E}.
\label{time}
\end{equation}
For massless particles, since they can have arbitrary small amount of energy
$\Delta E=\hbar c/\lambda$, being $\lambda$ the wavelenght of the
particle, the range of the respective force is infinite. This happens for
photons and gluons in both, the usual QED and QCD theories and also in the
present approach. However, from the point of view of an observer in our
space-time (unprimed one) the range of a gluon compared
with a photon of the same energy is shorter:
\begin{equation}
R_{\mbox{gluon}}=R_{\mbox{photon}}\cdot\frac{c'}{c}.
\label{range}
\end{equation}
As long as $c'\ll c$ we see that the range of the gluon as seeing in our
laboratories is rather smaller than the range of a photon with the same
energy. Hence, gluons would appear to us as being very massive.

Let us summarise the picture we have built up.
In our context, the confinement which can be
understood as: i)
there are no colored states in Nature or, ii) scattering of hadrons
produces only hadrons,
are equivalent by construction. This is so because the space-time dimensions
accessible to the ordinary electron-photon devices are only sensitive to the
electric part of the quarks. All electroweak degrees of freedom as
leptons, photons,
intermediate vector bosons and Higgs bosons, run on the usual $x$-world,
while gluons run only in the $x'$ world.
Only quarks are, in some sense,  the ``bridge''
of both type of space-time since they carry color and electric charge.
Since ``observation'' means that the system has experienced some interaction,
such as the scattering off it of light or electron, in the absence of
any interaction, the system is totally isolated from the outside world.
It is in this sense that the color confinement is understood in our
approach. According to the special theory of relativity, length and
time measurements are dependent on the observer. In our scheme those
measurements depend also on the type of fields used in carrying out
the measurements: photons or gluons.

In despite of its speculative nature,
our proposition can, at least in principle, be verified
experimentally.
Firstly, by observing colorless fractional charges.
Within our context, even if it is possible to produce
colored states, it is easy to see that it is not possible to detect,
directly, these degrees of freedom. This is because  all
the experimental apparatus used so far are electron-photon devices.
Hence, according
to our theory they are not sensitive to the primed world space-time.
Then, it is only throughout the electric interactions of
quarks that we have information about the hidden degrees of freedom.
Secondly, by evaluating the quark form factors $W_G$ and $W_A$ in
Eq~(\ref{def1}) and (\ref{def2}), respectivelly, using them for calculating
the nucleon (hadrons) form factors and verifying if they coincide or not with
the observed ones. Thirdly, looking for effects of non-renormalizable
interactions like that in Eq.~(\ref{yuca}) or its simplified form in
Eq.~(\ref{sim}).

It is still posible that the space-time coordinates transform under
$L\otimes L'$ like $({\bf4},{\bf4}')$.

Finally, we call attention to the fact that although we have extended the
number of space-time dimensions, any process like ``compactification'' of
the extra dimension is not necessary.
Thus, this ideas could be interesting in other
contexts not necessarily in the one we are concerned here.

\newpage

\acknowledgements

We would like to thank the
Con\-se\-lho Na\-cio\-nal de De\-sen\-vol\-vi\-men\-to Cien\-t\'\i
\-fi\-co e Tec\-no\-l\'o\-gi\-co (CNPq) for partial
financial support. I also thank R. Aldrovandi, C. O. Escobar, L. A. Ferreira
and G. Matsas for useful discussions.

\end{document}